# Electron and Hole Mobilities in Single-Layer $WSe_2$

Adrien Allain and Andras Kis*

*Electrical Engineering Institute, Ecole Polytechnique Federale de Lausanne (EPFL), CH-1015 Lausanne, Switzerland*

**Correspondence should be addressed to: Andras Kis, andras.kis@epfl.ch*

## ABSTRACT

Single-layer transition metal dichalcogenide (TMD) $WSe_2$ has recently attracted a lot of attention because it is a 2D semiconductor with a direct band-gap. Due to low doping levels it is intrinsic and shows ambipolar transport. This opens up the possibility to realize devices with the Fermi level located in valence band, where the spin/valley coupling is strong and leads to new and interesting physics. As a consequence of its intrinsically low doping, large Schottky barriers form between $WSe_2$ and metal contacts, which impede the injection of charges at low temperatures. Here, we report on the study of single-layer $WSe_2$ transistors with a polymer electrolyte gate (PEO:$LiClO_4$). Polymer electrolytes allow the charge carrier densities to be modulated to very high values, allowing the observation of both the electron- and the hole-doped regimes. Moreover, our ohmic contacts formed at low temperatures allow us to study the temperature dependence of electron and hole mobilities. At high electron densities, a re-entrant insulating regime is also observed, a feature which is absent at high hole densities.

Semiconducting transition metal dichalcogenides have recently emerged as promising materials for electronics[1] and optoelectronics,[2-5] exhibiting high charge carrier mobilities[6] and mechanical robustness.[7] These materials are particularly interesting in their monolayer form, where they exhibit a direct band-gap and a broken inversion symmetry, which results in a coupled spin and valley degree of freedom.[8-10] So far, $MoS_2$ has been the most widely studied TMD, due to its natural abundance. $MoS_2$ appears to be naturally n-doped, which has the advantage of allowing easy injection of charge carriers into the conduction band, but makes it difficult to achieve p-type transport which up to now was only observed in thicker samples.[11] On the other hand, monolayer $WSe_2$ has already been shown to exhibit ambipolar behavior,[12-14] however at the cost of difficulties in realizing good electrical contacts. Reaching the valence band is interesting from a technological standpoint, as it opens the possibility of designing planar p-n junctions,[3-5] but it also holds promises for spintronics, since large spin coherence time[8] and electrical control of the spin[15] have been predicted.

It is however difficult to engineer contacts to 2D semiconductors that perform satisfactorily at low temperatures. The low-temperature transport of naturally n-doped $MoS_2$ has been difficult to study due to the non-ohmic nature of electrical contacts. Top-gated geometries[6] or *in-situ* annealing[16] are required in order to achieve high carrier densities required for the electrodes to work properly. In intrinsic 2D semiconductors such as $WS_2$ and $WSe_2$, this problem is even more prominent. One possible solution is to chemically dope the contact area in order to induce very high carrier densities and produce ohmic behavior.[12, 14]

Here we report the study of single-layer $WSe_2$ transistors gated by the polymer electrolyte PEO:$LiClO_4$ (see Figure 1a). The superior electrostatic control offered by this ion gel allows very high carrier densities to be reached and similar electrolytes have previously allowed the observation of ambipolar behavior in $WS_2$ at room temperature.[17-18] Moreover, the atomic-scale thickness of the electric double layer (EDL) formed at the interface (see Figure 1b), which acts as a super-efficient capacitor, thins the Schottky barrier down to a regime where tunneling becomes the main charge injection mechanism.[19] This allows efficient injection of both types of charge carriers, even at cryogenic temperatures, and to perform, for the first time, the study of the transport properties of single-layer $WSe_2$ as a function of temperature and both p and n-type behavior on the same device. Figure 1a shows an optical micrograph of the finalized device. The device was annealed *in-situ* at 80°C in high vacuum prior to measurements.

## RESULTS AND DISCUSSION

Figure 1c describes the behavior of the transistor at room temperature. When compared to ionic liquids, such as the widely-used DEME-TFSI,[20-22] PEO:$LiClO_4$ has a slower response but also superior mechanical stability at low temperatures and against repeated temperature cycling. The lower ionic conductivity of PEO:$LiClO_4$ is partially compensated by its large chemical stability window. At $T = 300$ K, polymer electrolyte gate voltages ($V_{PE}$) as low as $V_{PE} = -4$ V can be applied with minimum leakage current $I_{leak} < 200$ pA. We can identify three regimes of interest (ROI) that we will focus on in the rest of this letter: the hole transport regime (labeled "*p*-side"), the electron transport regime ("*n*-side") and a regime at high electron density where the conductivity saturates ("high-n regime"). After the measurements as a function of temperature in these three regimes were completed, another transfer curve was acquired at a slightly higher temperature ($T = 320$ K, see Figure 1d). We can see that it results in significantly higher conductivities. A careful analysis also reveals that the apparent bandgap is reduced. Similarly to previous observations when using DEME-TFSI on ZnO,[22] we speculate that this effect is due to the higher carrier densities that can be accumulated at higher temperature for a given $V_{PE}$. For the low-temperature study, we proceeded in the following way: we apply a given value of $V_{PE}$ for several hours at $T = 300$ K and wait for the current to stabilize. In this way we make sure that all the mobile ions in our gel had enough time to migrate to the EDL. This typically results in a substantial but reversible shift of the entire transfer curve, if acquired afterwards (compare for example Figures 1c and 1d). The charge state of the sample could however be restored by baking it *in-situ* to 80°C for several hours, indicating that no permanent modification/degradation had occurred. We performed this charge "reset" before the study of each ROI in temperature. Finally, the ionic conductivity of PEO:$LiClO_4$ rapidly drops as the temperature is lowered and the ions freeze around $T = 270$ K. In order to study the transport and field-effect mobility of the devices, we used the back-gate voltage $V_G$ to modulate the charge carrier density below the freezing point. Our sample was remarkably stable below the freezing point and the back-gate voltage or temperature history did not influence the measurements, as can be seen on Figure 2, where voltage traces and retraces are shown to exhibit very little hysteresis.

Figure 2 demonstrates that the use of the ion gel substantially decreases the contact resistance at cryogenic temperatures, thus allowing proper 4-point measurements of the conductivity to be performed. Figure 2a (for the "*p*-side") and Figure 2b (for the "*n*-side") show the back-gate dependence at $T = 4$ K of the 4-point resistance $R_{4P}$, 2-point resistance $R_{2P}$, and the inferred contact resistance:

$$R_{contacts} = R_{2P} - \frac{L_{tot}}{L_{xx}} R_{4P}$$

Where $L_{tot} = 5.3\ \mu\text{m}$ and $L_{xx} = 2.4\ \mu\text{m}$ are the total source-drain length and distance separating the voltage side-probes, respectively (see Figure 1a). The contact resistance decreases with increasing charge carrier concentration and seems to asymptotically approach a finite value, as indicated by the dashed lines. These values corresponds to $\rho_c \simeq 10 - 13\ \text{k}\Omega \cdot \mu\text{m}$, a value which is lower than the best value recorded for single-layer $MoS_2$ transistors with solid top–gates.[6] The output characteristic $I(V)$ is also nearly ohmic, as can be seen in the inset of Figure 2a.

Now that we have demonstrated that good electrical contacts have been realized, we can perform a temperature study of the 4-point conductivity. Figure 3a (Figure 3b) shows the conductance as a function of back-gate voltage for different temperatures in the hole-doped (electron-doped) region. These two regimes show distinct behaviors. On the *n*-side, the behavior is very similar to previously studied electron transport in $MoS_2$. The conductance increases with decreasing temperature, indicating a metallic behavior. We can measure the carrier density using the Hall effect (inset of Figure 3d) and calculate the field-effect and Hall mobilities. The field-effect mobility, reported in Figure 3b for holes and Figure 3d for electrons, is given by:

$$\mu_{FE} = \frac{1}{C_G} \frac{L_{xx}}{W} \frac{\partial G}{\partial V_G}$$

Where $W = 2.9\ \mu\text{m}$ and $L_{xx} = 2.4\ \mu\text{m}$ are the width and the length of the device, $G$ is the conductance and $C_G = 1.39 \times 10^{-8}\ \text{F} \cdot \text{m}^{-2}$ is the back-gate capacitance inferred from the Hall effect (see inset of Figure 3d). It starts at around 30 $\text{cm}^2\text{V}^{-1}\text{s}^{-1}$ at room temperature and reaches a value of 160 $\text{cm}^2\text{V}^{-1}\text{s}^{-1}$ at $T = 4$ K. The Hall mobility at $T = 4$ K (see Figure S1), which is proportional to the drift mobility of the electrons, is given by:

$$\mu_{Hall} = \frac{\sigma}{en}$$

Where $\sigma$ is the conductivity, $e$ the electron charge and $n$ the electron density extracted from the Hall effect measurements. Similarly to the case of $MoS_2$,[16] it is dependent on the carrier density and asymptotically approaches the value of the field-effect mobility at high carrier densities. It reaches a value of 100 $\text{cm}^2\text{V}^{-1}\text{s}^{-1}$ at $T = 4$ K for $n = 1.9 \times 10^{13}\ \text{cm}^{-2}$. The p-side differs from the n-side in several respects: it is non-metallic, has a higher and almost temperature-independent field-effect mobility and the Hall effect could not be measured.

As shown on Figure 3a, the *p*-side displays values of conductance that are similar to those recorded on the *n*-side near room temperature. The temperature dependence is however different, with the conductivity slowly decreasing with decreasing temperature thus displaying an insulating behavior. It seems that a metal-insulator transition can be extrapolated from the conductance curves in Figure 3c. Temperature scans performed at a later time and higher hole doping levels revealed that a metallic regime can indeed be reached on the *p*-side (see Figure S2), associated with perfectly ohmic (linear) *I(V)* curves (see Figure S3). Contrary to the *n*-side and to previous results in $MoS_2$,[6] the metal-insulator transition happens at a value of the sheet conductivity $\sigma \approx 7.5\, e^2/h$, which is much higher than the value predicted by the Ioffe-Regel criterion.[23] Inhomogeneous current density (due to flake damage for example) can be ruled out because the conductance measured from both sides of the flake is the same (see Figure S4). This could instead be due to a decrease in the density of thermally excited carriers with decreasing temperature, or to a change of the bandgap with temperature, all leading to a non-constant carrier density at fixed back-gate voltage. This effect could be more pronounced on the *p*-side because the Fermi level naturally lies closer to the conduction band (see Figure 1c). The conductance nevertheless displays a linear regime and the field-effect mobility for holes can be derived (see Figure 3c). It is systematically higher than the electron mobility, with a very weak temperature dependence. This indicates that the mobility is not yet phonon-limited and is limited by intrinsic sources of scattering. The room-temperature ($T = 250$ K) field-effect mobility of holes in $WSe_2$ (180 $cm^2V^{-1}s^{-1}$) is significantly higher than the best field-effect electron mobilities reported in $MoS_2$ (60-100 $cm^2V^{-1}s^{-1}$) and consistent with previously reported values for $WSe_2$ (refs [12] and [13]).

Compared to the hole mobility, the electron mobility starts to drop drastically above $T = 100$ K (see Figure 3d) and reaches 30 $cm^2.V^{-1}.s^{-1}$ at room temperature. This value is lower than previous reports of room temperature electron mobility in SL $WSe_2$.[13-14] We also notice that this mobility drop is concomitant with the onset of a sublinear regime in the conductance *vs.* back-gate voltage. Looking back at Figure 1c, this might be the onset of the conductance saturation observed at high electron densities. Such saturation/decrease of the conductivity at high carrier densities (see Figure S5 for another example of the conductance drop at high positive $V_{PE}$ at room temperature in another device) have been reported in bi-layer and tri-layer graphene and attributed to the onset of interband scattering.[21] We know that the band structure of TMDs is affected by quantum confinement in the 2D limit.[24] Increased interband scattering could be one of the reasons why the mobility in single-layers is systematically lower than that of bulk material.[25] Interestingly, the high-temperature dependence of the electron mobility in $MoS_2$ is very similar,[6] which points to a similar scattering mechanism being at the origin of the mobility reduction.

Figure 4 shows the temperature and back-gate voltage dependence of the conductance, taken at high electron doping. Interestingly, the behavior is insulating, with a dependence on gate voltage that reverts as the temperature is decreased. This re-entrant insulating regime at high carrier densities is unexpected and not understood. It is not related to sample damage, as the behavior was fully reversible after warming up the sample.

## CONCLUSION

In summary, we carried out the first study of the electron and hole transport in single-layer $WSe_2$ transistors as a function of temperature. The use of a polymer electrolyte gate allowed us to achieve ohmic contacts at cryogenic temperatures, which is a pre-requisite for any electrical measurement. The temperature study of the electron-doped side reveals a behavior very similar to that of single-layer $MoS_2$, with a metallic state above $n = 8 \times 10^{12}$ $cm^{-2}$ and a strong suppression of the mobility at high temperatures. The hole-doped region shows a metal-insulator transition at much higher conductivity then expected from the Ioffe-Regel criterion. This speaks for a strong temperature dependence of the carrier density. The field-effect mobility is nearly temperature-independent, with a room-temperature value which is consistent with previously-reported values[12] and significantly higher than that of electrons. The regime with high levels of electron doping is surprisingly insulating. This work demonstrates for the first time that ohmic contacts can be formed with $WSe_2$ at low temperatures and is the first report of a temperature-depended hole transport in a two-dimensional TMD. It thus paves the way for the realization of electrical devices taking advantage of the peculiar structure of the valence band of TMDs and to the demonstration of spin/valleytronic devices using these materials.

## METHODS

### Device fabrication

$WSe_2$ monolayer flakes were exfoliated onto $Si/SiO_2$ wafers and identified from their optical contrast.[26] Standard e-beam lithography and thermal evaporation were employed to create electrical contacts. These consist of 5 nm of Pd,and 45 nm of Au and a capping layer of 15 nm of $SiO_2$ in order to isolate the metal contacts from the ionic gel. The gate electrode is made of 30 nm of Pt and was deposited during a separate evaporation. Its surface area is much larger than that of the $WSe_2$ flake, ensuring that most of the applied voltage drops at the surface of the $WSe_2$ flake, maximizing the gate efficiency. $WSe_2$ flakes were shaped into Hall bars with a well-defined channel geometry using oxygen plasma etching.

### Electrolyte Preparation

The polymer electrolyte was prepared by dissolving an 8:1 mixture of poly-ethylene oxide (PEO) and lithium perchlorate ($LiClO_4$)[27] in acetonitrile to form a 10 wt% solution. The solution was sonicated and filtered using a syringe filter.[28] A droplet was deposited onto the device inside a glovebox (Ar atmosphere) and left to dry. The device had previously been degased for four hours at 100°C in the glovebox to remove adsorbates.

## ACKNOWLEDGEMENTS

The authors would like to thank D. Lembke for his help in preparing the ion gels and T. Heine and A. Morpurgo for useful discussions. Device fabrication was carried out in part in the EPFL Center for Micro/Nanotechnology (CMI). We thank Z. Benes (CMI) for technical support with e-beam lithography.

## REFERENCES

1. Radisavljevic, B.; Radenovic, A.; Brivio, J.; Giacometti, V.; Kis, A., Single-Layer $MoS_2$ Transistors. *Nat. Nanotechnol.* **2011,** 6, 147-150.
2. Lopez-Sanchez, O.; Lembke, D.; Kayci, M.; Radenovic, A.; Kis, A., Ultrasensitive Photodetectors Based on Monolayer $MoS_2$. *Nat. Nanotechnol.* **2013,** 8, 497-501.
3. Baugher, B. W. H.; Churchill, H. O. H.; Yang, Y.; Jarillo-Herrero, P., Optoelectronic Devices Based on Electrically Tunable P-N Diodes in a Monolayer Dichalcogenide. *Nat. Nanotechnol.* **2014,** 9, 262-267.
4. Pospischil, A.; Furchi, M. M.; Mueller, T., Solar-Energy Conversion and Light Emission in an Atomic Monolayer P-N Diode. *Nat. Nanotechnol.* **2014,** 9.
5. Ross, J. S.; Klement, P.; Jones, A. M.; Ghimire, N. J.; Yan, J.; Mandrus, D. G.; Taniguchi, T.; Watanabe, K.; Kitamura, K.; Yao, W.*, et al.*, Electrically Tunable Excitonic Light-Emitting Diodes Based on Monolayer $WSe_2$ P-N Junctions. *Nat. Nanotechnol.* **2014,** 9, 268-272.
6. Radisavljevic, B.; Kis, A., Mobility Engineering and a Metal–Insulator Transition in Monolayer $MoS_2$. *Nat. Mat.* **2013,** 12, 815-820.
7. Bertolazzi, S.; Brivio, J.; Kis, A., Stretching and Breaking of Ultrathin $MoS_2$. *ACS Nano* **2011,** 5, 9703-9709.
8. Xiao, D.; Liu, G.-B.; Feng, W.; Xu, X.; Yao, W., Coupled Spin and Valley Physics in Monolayers of $Mos_2$ and Other Group-Vi Dichalcogenides. *Phys. Rev. Lett.* **2012,** 108, 196802.
9. Mak, K. F.; He, K.; Shan, J.; Heinz, T. F., Control of Valley Polarization in Monolayer $MoS_2$ by Optical Helicity. *Nat. Nanotechnol.* **2012,** 7, 494–498.
10. Zeng, H.; Dai, J.; Yao, W.; Xiao, D.; Cui, X., Valley Polarization in Mos2 Monolayers by Optical Pumping. *Nat. Nanotechnol.* **2012,** 7, 490.
11. Zhang, Y.; Ye, J. T.; Matsuhashi, Y.; Iwasa, Y., Ambipolar $MoS_2$ Thin Flake Transistors. *Nano Lett.* **2012**.
12. Fang, H.; Chuang, S.; Chang, T. C.; Takei, K.; Takahashi, T.; Javey, A., High-Performance Single Layered $WSe_2$ P-Fets with Chemically Doped Contacts. *Nano Lett.* **2012,** 12, 3788-3792.
13. Liu, W.; Kang, J.; Sarkar, D.; Khatami, Y.; Jena, D.; Banerjee, K., Role of Metal Contacts in Designing High-Performance Monolayer N-Type $WSe_2$ Field Effect Transistors. *Nano Lett.* **2013,** 13, 1983-1990.
14. Fang, H.; Tosun, M.; Seol, G.; Chang, T. C.; Takei, K.; Guo, J.; Javey, A., Degenerate N-Doping of Few-Layer Transition Metal Dichalcogenides by Potassium. *Nano Lett.* **2013,** 13, 1991-1995.
15. Gong, K.; Zhang, L.; Liu, D.; Liu, L.; Zhu, Y.; Zhao, Y.; Guo, H., Electric Control of Spin in Monolayer $WSe_2$ Field Effect Transistors. *arXiv:1310.1816 [cond-mat]* **2013**.
16. Baugher, B. W. H.; Churchill, H. O. H.; Yang, Y.; Jarillo-Herrero, P., Intrinsic Electronic Transport Properties of High-Quality Monolayer and Bilayer $MoS_2$. *Nano Lett.* **2013,** 13, 4212-4216.
17. Braga, D.; Gutiérrez Lezama, I.; Berger, H.; Morpurgo, A. F., Quantitative Determination of the Band Gap of $WS_2$ with Ambipolar Ionic Liquid-Gated Transistors. *Nano Lett.* **2012**.
18. Jo, S.; Ubrig, N.; Berger, H.; Kuzmenko, A. B.; Morpurgo, A. F., Mono- and Bilayer $WS_2$ Light-Emitting Transistors. *Nano Lett.* **2014**.

19. Perera, M. M.; Lin, M.-W.; Chuang, H.-J.; Chamlagain, B. P.; Wang, C.; Tan, X.; Cheng, M. M.-C.; Tománek, D.; Zhou, Z., Improved Carrier Mobility in Few-Layer $MoS_2$ Field-Effect Transistors with Ionic-Liquid Gating. *ACS Nano* **2013,** 7, 4449-4458.
20. Ye, J. T.; Zhang, Y. J.; Akashi, R.; Bahramy, M. S.; Arita, R.; Iwasa, Y., Superconducting Dome in a Gate-Tuned Band Insulator. *Science* **2012,** 338, 1193-1196.
21. Ye, J.; Craciun, M. F.; Koshino, M.; Russo, S.; Inoue, S.; Yuan, H.; Shimotani, H.; Morpurgo, A. F.; Iwasa, Y., Accessing the Transport Properties of Graphene and Its Multilayers at High Carrier Density. *Proc. Natl. Acad. Sci. U. S. A.* **2011,** 108, 13002-13006.
22. Yuan, H.; Shimotani, H.; Tsukazaki, A.; Ohtomo, A.; Kawasaki, M.; Iwasa, Y., High-Density Carrier Accumulation in ZnO Field-Effect Transistors Gated by Electric Double Layers of Ionic Liquids. *Adv. Funct. Mater.* **2009,** 19, 1046–1053.
23. Lee, P. A.; Ramakrishnan, T. V., Disordered Electronic Systems. *Rev. Mod. Phys.* **1985,** 57, 287-337.
24. Kumar, A.; Ahluwalia, P. K., Electronic Structure of Transition Metal Dichalcogenides Monolayers 1h-Mx2 (M = Mo, W; X = S, Se, Te) from *ab-initio* Theory: New Direct Band Gap Semiconductors. *Eur. Phys. J. B* **2012,** 85, 1-7.
25. Fivaz, R.; Mooser, E., Mobility of Charge Carriers in Semiconducting Layer Structures. *PhRv* **1967,** 163, 743-755.
26. Benameur, M. M.; Radisavljevic, B.; Heron, J. S.; Sahoo, S.; Berger, H.; Kis, A., Visibility of Dichalcogenide Nanolayers. *Nanotechnology* **2011,** 22, 125706.
27. Kruger, M.; Buitelaar, M. R.; Nussbaumer, T.; Schonenberger, C.; Forro, L., Electrochemical Carbon Nanotube Field-Effect Transistor. *Appl. Phys. Lett.* **2001,** 78, 1291-1293.
28. Dankerl, M.; Tosun, M.; Stutzmann, M.; Garrido, J. A., Solid Polyelectrolyte-Gated Surface Conductive Diamond Field Effect Transistors. *Appl. Phys. Lett.* **2012,** 100, 023510.

# FIGURES

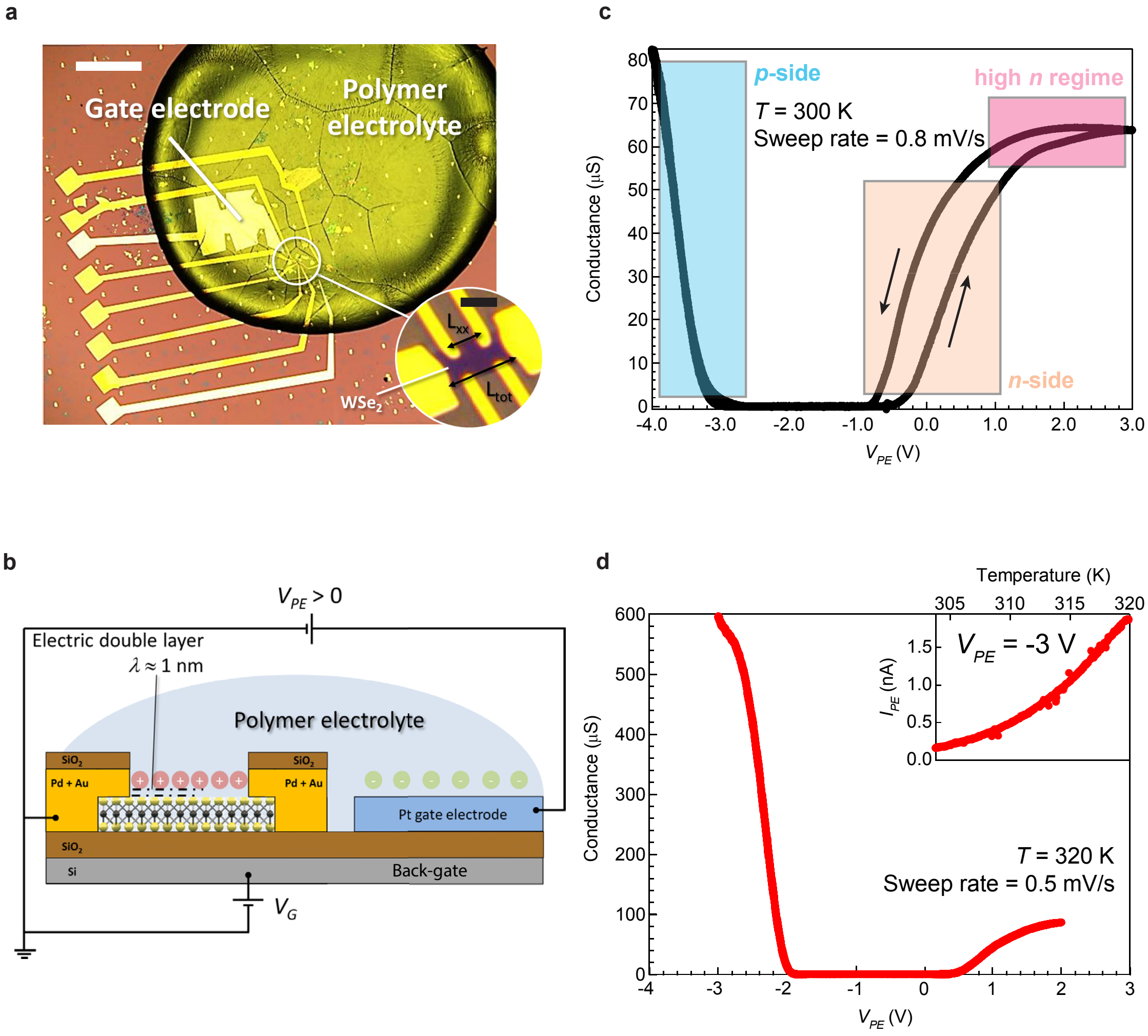

**Figure 1. Polymer electrolyte-gated $WSe_2$ single-layer transistors**. **a**, Optical micrograph of the device. The scale bar is 500 µm long. Inset: zoom on the transistor region. Scale bar is 3 µm long. **b**, Sketch of the charge configuration in the device for a positive voltage $V_{PE}$ applied to the polymer electrolyte gate electrode. λ is the Debye length. **c**, Room-temperature (T = 300 K) transfer curve of the transistor. We can identify three regions of interest: hole-doped (hereafter referred to as “p-side”), slightly electron-doped (“n-side”) and highly electron-doped (“high n regime”). **d**, High-temperature (T = 30 K) transfer curve of the same device. Inset: leakage current at $V_{PE}$ = -3 V as a function of temperature.

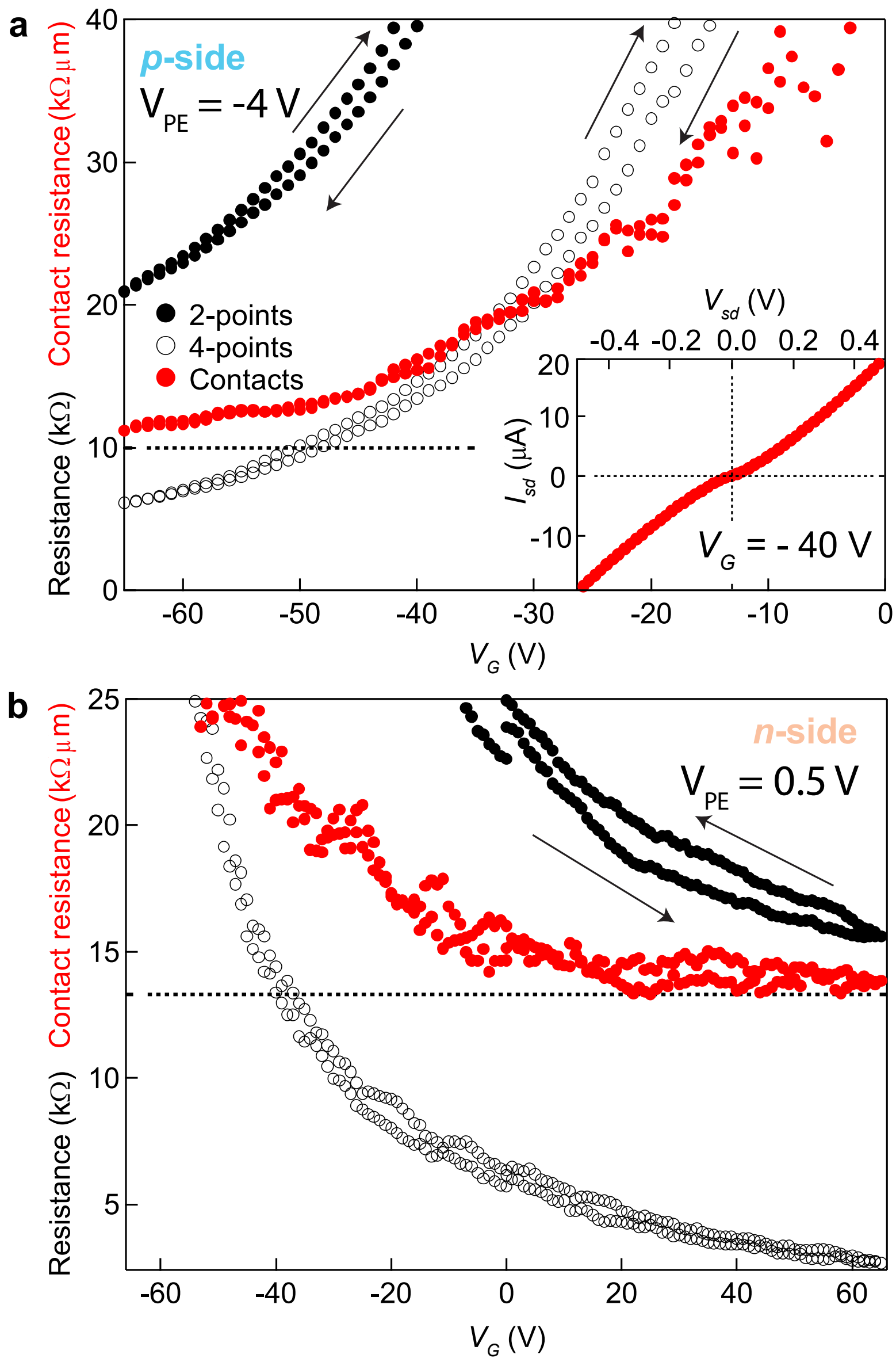

**Figure 2. Contact resistance at low temperature (*T* = 4K).** Back-gate dependence of the 2-point- (filled black dots), 4-point- (open black dots) and contact resistance (red dots) **a,** On the hole side. **b,** On the electron side. The dotted line is a guide to the eye representing the contact resistance achieved at high doping. The inset of (a) shows the *I*(*V*) output curve recorded at $V_G$ = -40V.

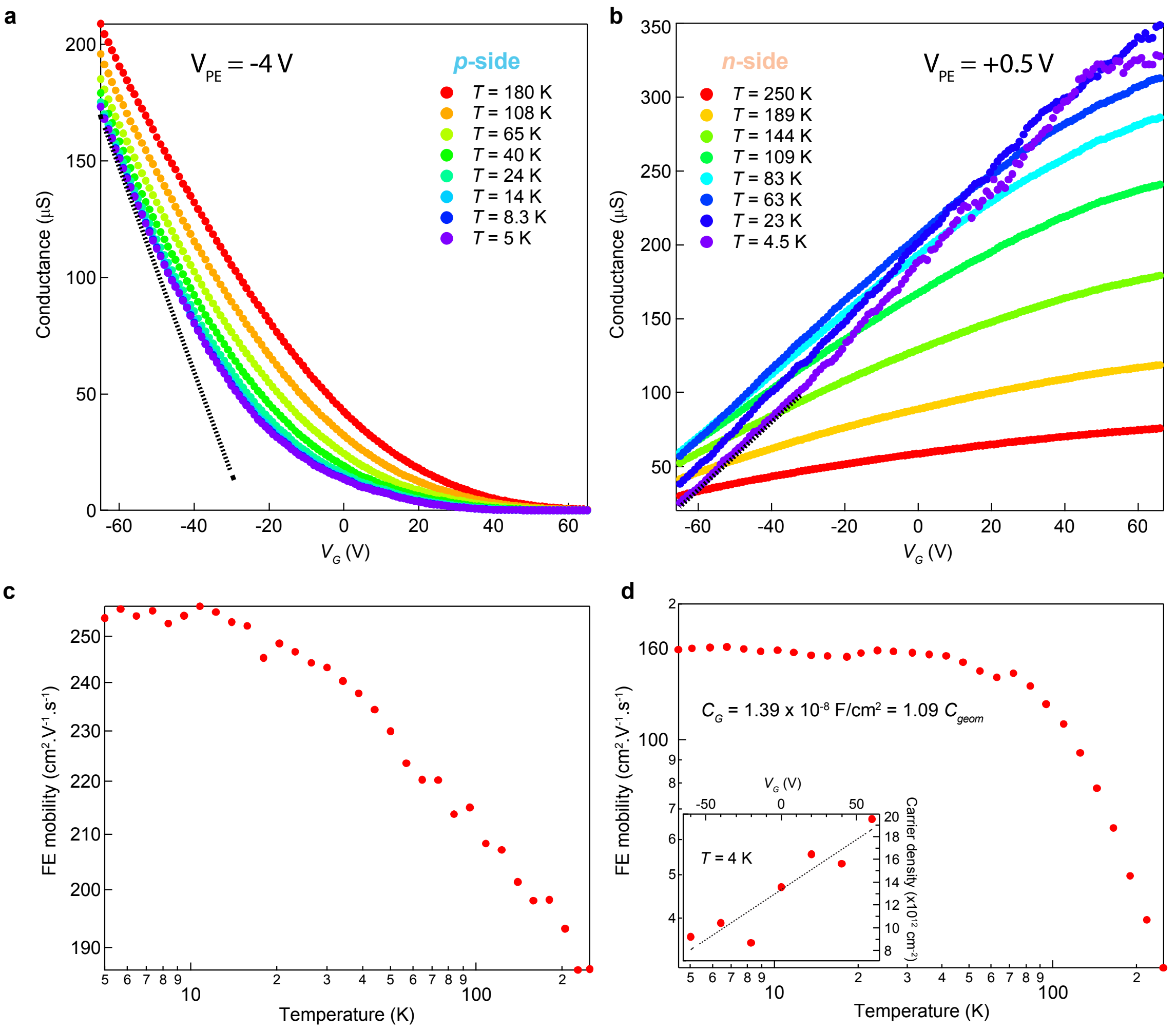

**Figure 3. Electron and holes mobilities in single-layer $WSe_2$.** Conductance as a function of back-gate voltage at different temperatures on the **a,** hole side and **b,** electron side. The corresponding field-effect mobilities as a function of temperature, on the *p*- **c,** and *n*-side **d**. In calculating the mobilities, we used the values of $\partial G/\partial V_G$ obtained in regions where the conductance displays a linear dependence on gate voltage (dashed lines in Figure 3a and Figure 3b). Inset of **d**: carrier density on the *n*-side, measured using the Hall effect. The extracted back-gate capacitance is indicated.

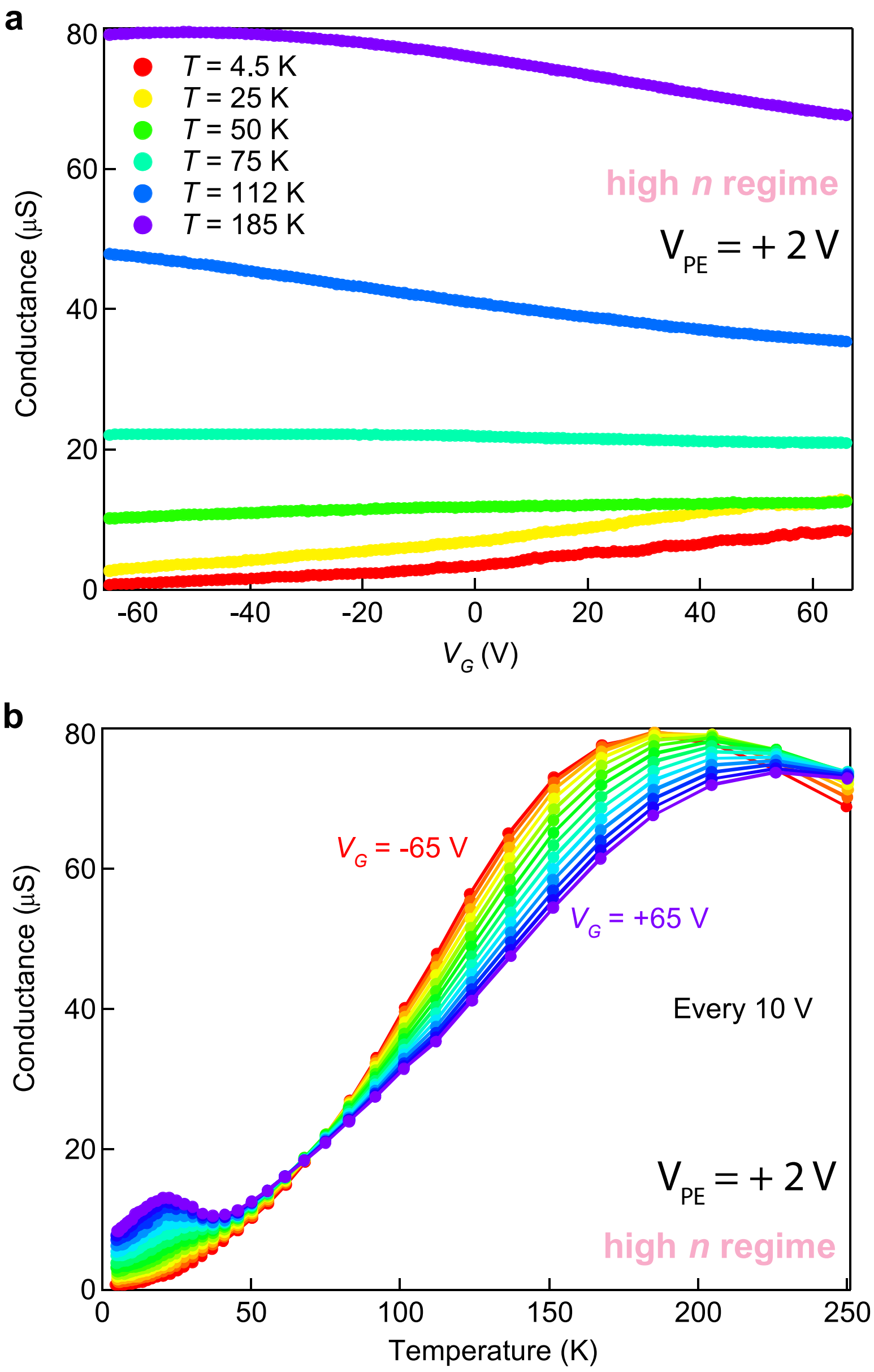

**Figure 4. Insulating behavior in the highly electron-doped regime. a,** Conductance *vs.* back-gate voltage at different temperatures. **b,** Conductance *vs.* temperature for different back-gate voltages.

# Supporting Material

# for

# "Electron and Hole Mobilities in Single-Layer $WSe_2$"

Adrien Allain and Andras Kis*

*Electrical Engineering Institute, Ecole Polytechnique Federale de Lausanne (EPFL), CH-1015 Lausanne, Switzerland*

**Correspondence should be addressed to: Andras Kis, andras.kis@epfl.ch*

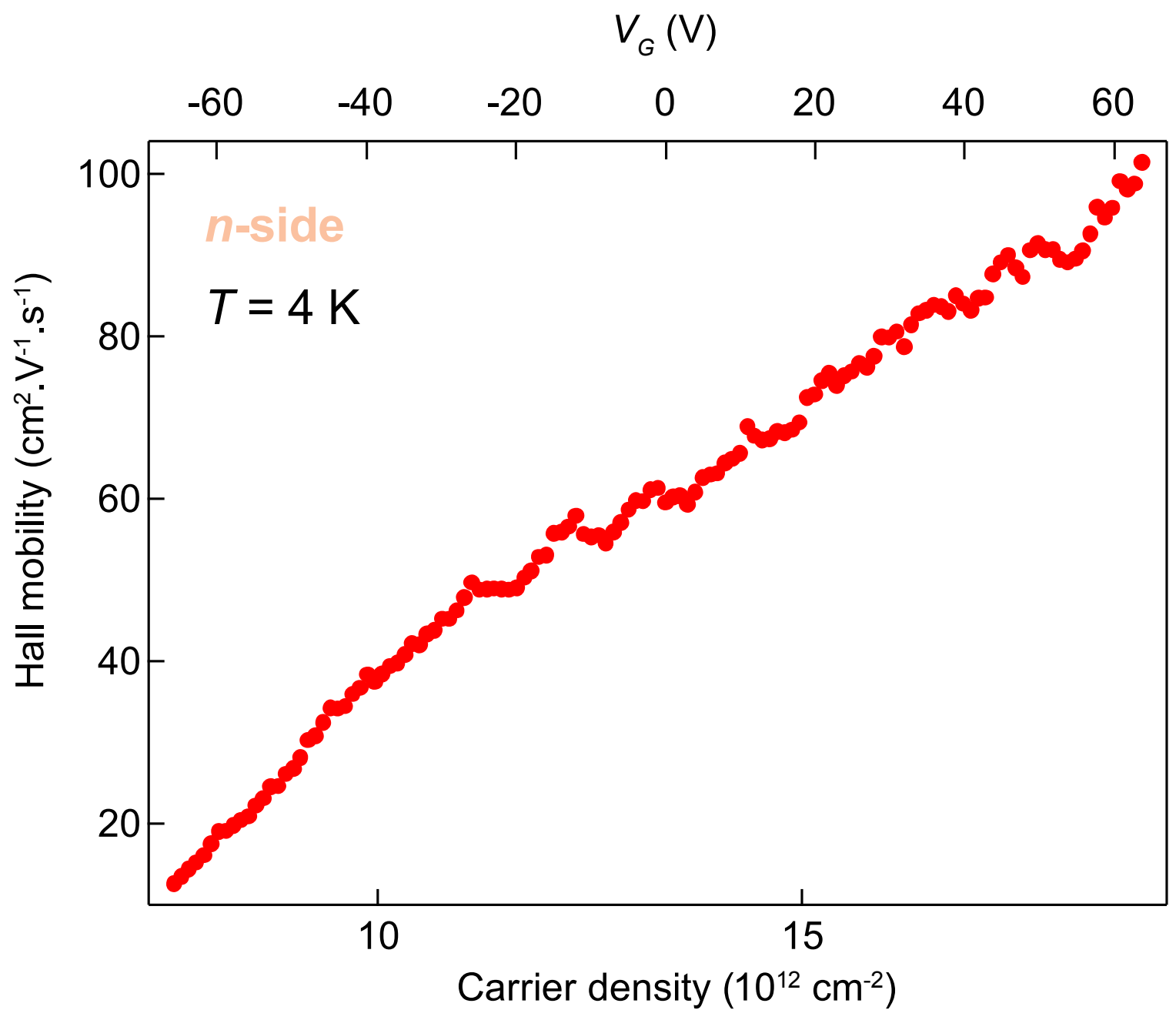

**Figure S1. Hall mobility of electrons.** The carrier density was extracted by measuring the Hall effect for several values of the back-gate voltage (Figure 3d in the main text). The carrier density could then be extrapolated to any back-gate voltage value using a linear fit. This allowed us to calculate the Hall mobility as a function of back-gate voltage/carrier density.

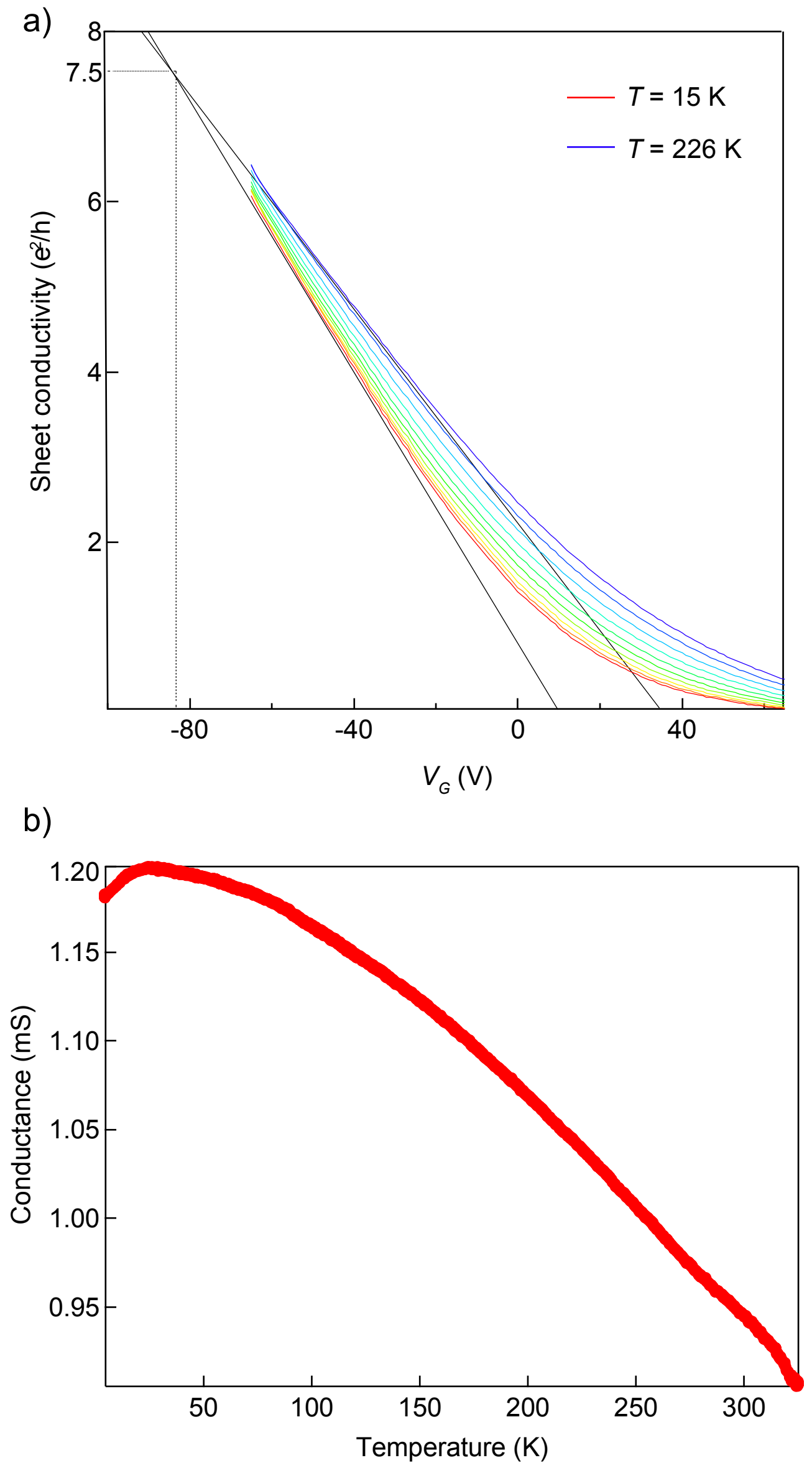

**Figure S2. Metal-insulator transition on the p-side.** (a) Back-gate dependence of the sheet conductivity (in units of the quantum of conductance) obtained during a subsequent cooling on the p-side ($V_{PE}$ = -4V). More time was given to the ions to migrate andthe doping was slightly higher than during the first p-side run and closer to the metal-insulator transition. Extrapolating the conductance curves, one can give an estimate of its position. (b) At a later stage of the experiment, the polymer-electrolyte voltage was set to $V_{\mathrm{PE}}$ = -3 V at a temperature of T = 320 K. As can be seen on Figure 1d in the main text, this results in significantly higher conductance, presumably because higher carrier concentrations can be achieved. When cooled down from this point, the sample displays metallic behavior.

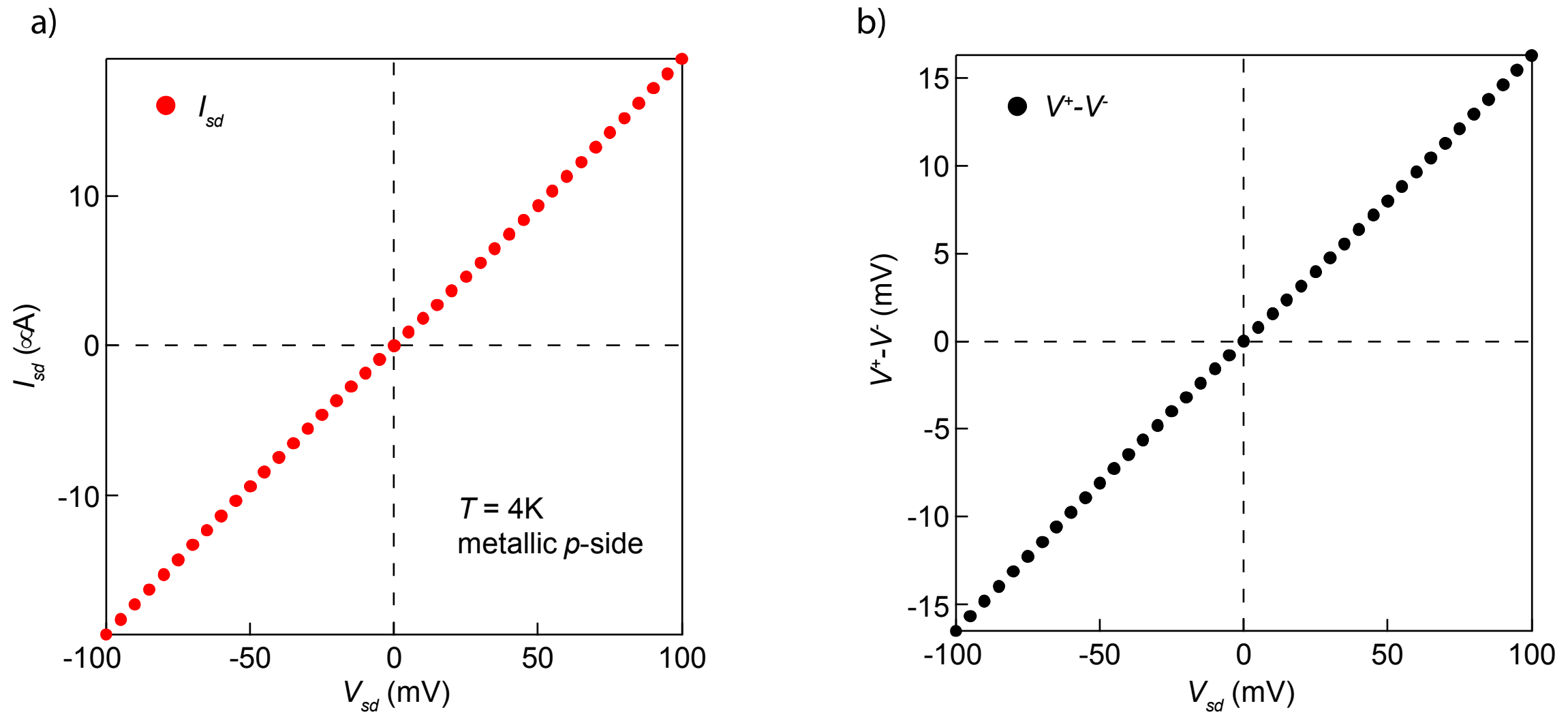

**Figure S3. Ohmic contacts at high hole doping. a)** *I(V)* output curve and b) Voltage drop between the side contacts in the highly hole-doped regime (measured after the cooldown from $V_{PE}$ = -3 V and $T$ = 320 K shown in Figure S2b). The device shows perfectly ohmic (linear) behavior at cryogenic temperatures.

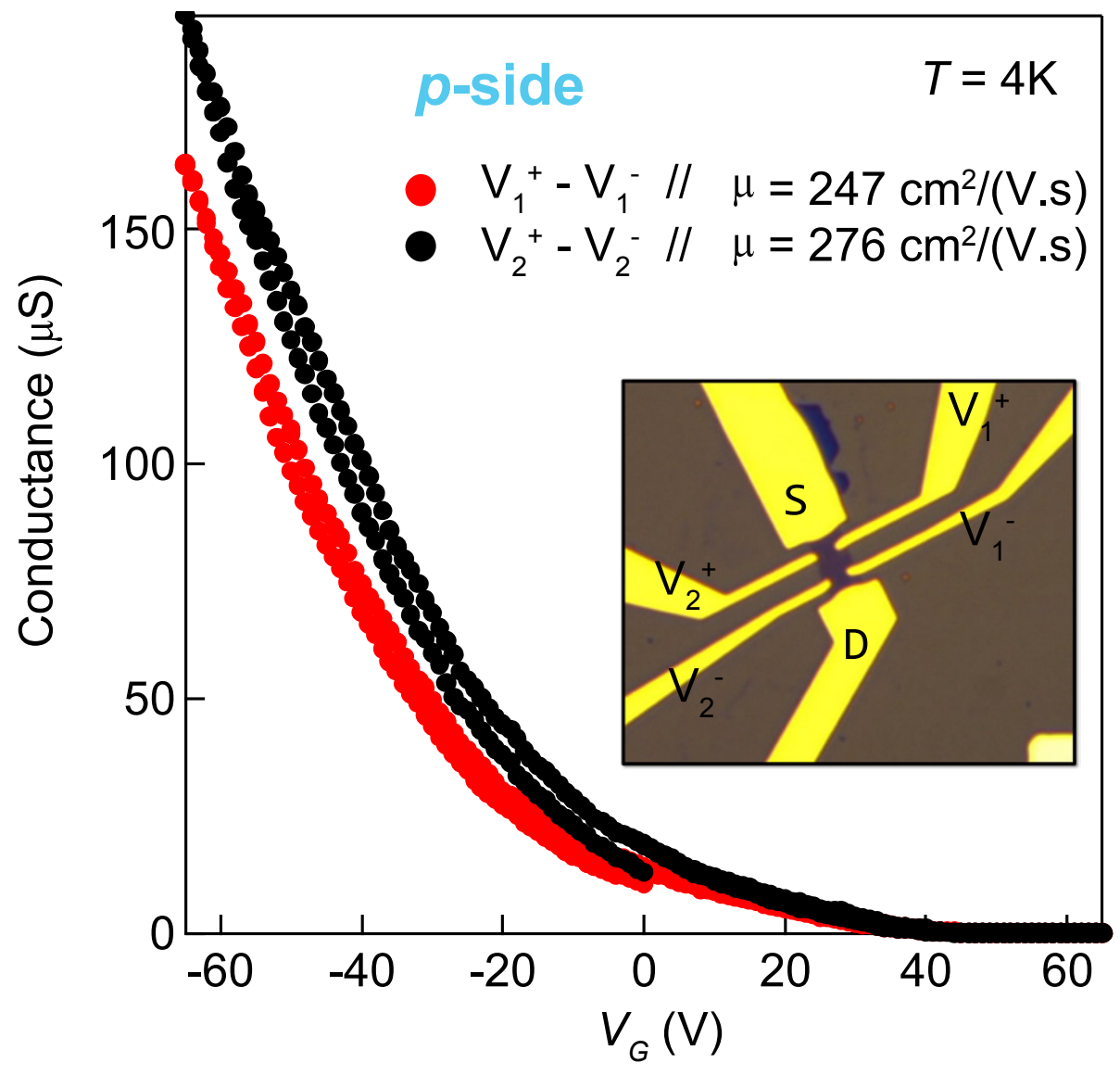

**Figure S4. Integrity of the flake.** The 4-point conductance measured from both sides of the flake gives almost the same result, which indicates that the physical integrity of the flake has been preserved.

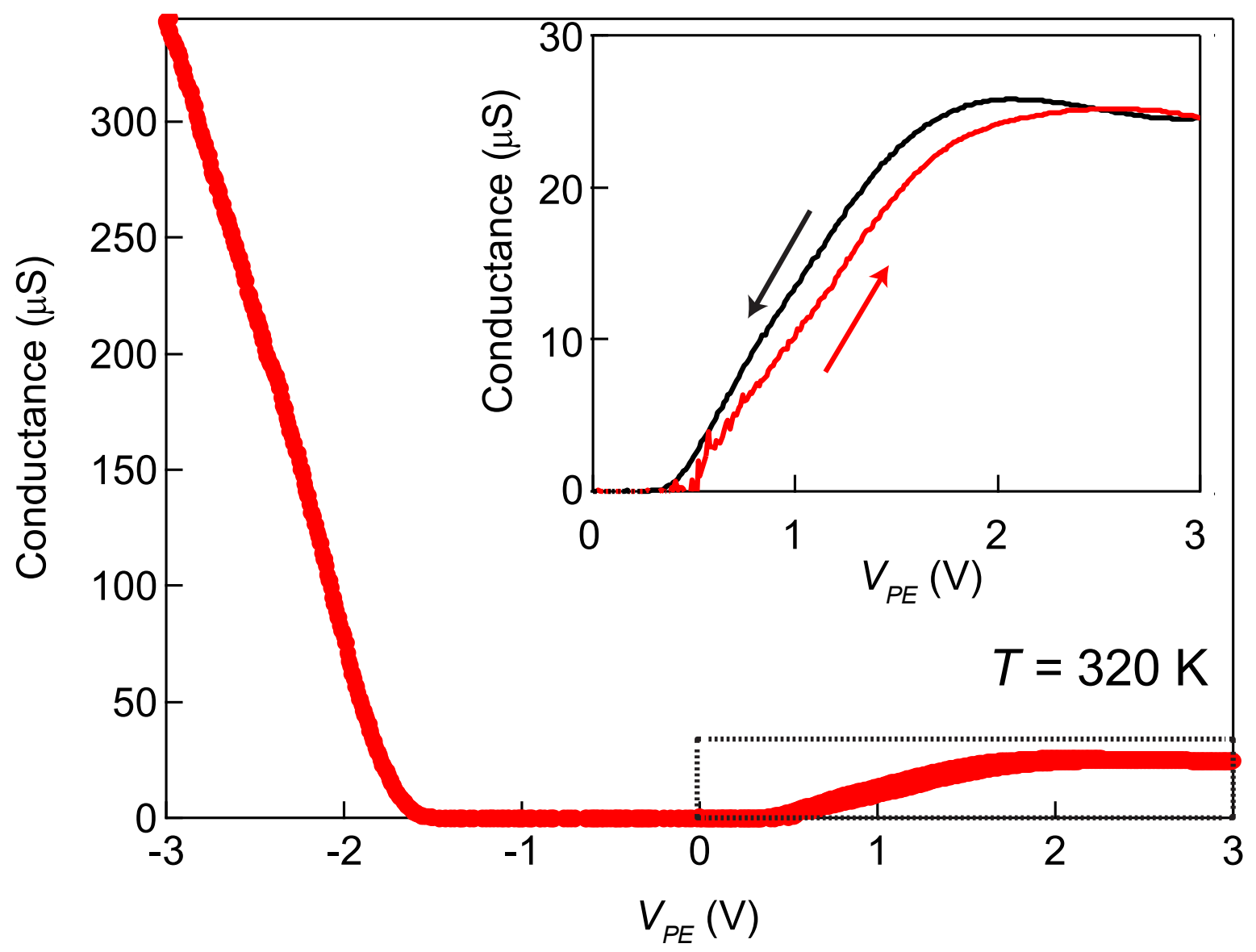

**Figure S5. Conductance saturation in the highly electron-doped region.** Conductance as a function of the polymer electrolyte gate voltage at $T$ = 320 K measured in another monolayer $WSe_2$ sample. Increasing the voltage to ±3 V reveals the strong asymmetry between electron and hole transport, with the conductance exhibiting a plateau and starting to decrease at high electron densities. Inset: zoom on the boxed region.